\documentclass[twocolumn,aps,prb,epsf]{revtex4}
\usepackage{amsmath}
\usepackage{epsfig}
\usepackage{epsf}
\usepackage{array}

\begin{document}

\title{Fictive Impurity Models: an Alternative Formulation of the Cluster
Dynamical Mean Field Method}
\author{S. Okamoto$^{(1)}$, A. J. Millis$^{(1)}$, H. Monien$^{(2)}$ and A.
Fuhrmann$^{(2)}$.}
\affiliation{$^{(1)}$Department of Physics, Columbia University\\
538 W. 120th St, NY NY 10027\\
$^{(2)}$Physikalisches Institut, Universitat Bonn\\
53115 Bonn, Germany}
\date{\today}

\begin{abstract}
`Cluster' extensions of the dynamical mean field method to include longer
range correlations are discussed. It is argued that the clusters arising in
these methods are naturally interpreted not as actual subunits of a physical
lattice but as algorithms for computing coefficients in an orthogonal
function expansion of the momentum dependence of the electronic self-energy.
The difficulties with causality which have been found to plague cluster
dynamical mean field methods are shown to be related to the `ringing'
phenomenon familiar from Fourier analysis. The analogy is used to motivate
proposals for simple filtering methods to circumvent them. The formalism is
tested by comparison to low order perturbative calculations and self
consistent solutions.
\end{abstract}

\maketitle



\section{Introduction}

Over the last decade the `dynamical mean field' method \cite{Georges96} has
emerged from earlier investigations\cite{Metzner88} of the infinite dimensional 
limit as a very useful tool for theoretical investigation of correlated electron
systems. It provides (within a certain approximation) a nonperturbative
means of obtaining the electron self-energy and spectral function, and
allows treatment of inelastic and thermal effects on the same footing as
ground state energetics. It has revealed new insights into the physics of
the Mott transition,\cite{Rozenberg95} of `colossal' magnetoresistance
manganites,\cite{Millis96b} of plutonium \cite{Savrasov01a} and many other
systems, and may be combined with LMTO band theory (for reviews see Refs.%
\onlinecite{Savrasov01b,Held01}) to treat correlation effects in realistic
models.

As originally formulated \cite{Georges96} the dynamical mean field method is
a \textit{local} approximation. The problem which one actually solves is a
single-site model self-consistently embedded in a medium. The results are an
exact representation of the physics of lattice models only in a limit of
infinite coordination number.\cite{Georges96,Metzner88} While the method
captures local physics very well, the treatment of intersite correlations is
an important open issue. Attempts to formulate a controlled expansion about
the infinite coordination limit have not led to useful and tractable
expressions. An alternative approach focuses on the self-consistent
embedding of a larger cluster in a medium,\cite%
{Hettler98,Moukouri00,Lichtenstein00,Kotliar02,Biroli02,Bolech02,Aryanpour03}
and is closely related to attempts to extend the `single-site CPA'
approximation to more than one site.\cite{DuCastelle75,Soloviev02} Most of
the approaches published to date involve choosing a specific cluster (subset
of sites of the actual lattice of interest) as well as a specific embedding
(geometry of connection of cluster sites to sites of the medium). An
alternative (`E-DMFT') method  involves using hybrid boson-fermion methods
to treat two-particle intersite correlations \cite{Si96,Chitra01,Sun02} and
has had some success.

While these approaches have led to a number of interesting results (for
recent examples see e.g. Refs.~\onlinecite{Lichtenstein00,Mazurenko02,Maier02}), one
worries that the choice of specific geometry both of the cluster and its
embedding in the medium may bias the physics. Further, the
real-space-cluster-based methods sometimes lead to self energies which are
non-causal for some momenta and frequencies. This is regarded as a grave
defect of the method and its cure is an important open problem.

In this paper we put forward an alternative point of view on the question of
extending dynamical mean field theory to include intersite correlations. Our
motivation is as follows. The original local dynamical mean field method,
although often described in terms of a site self consistently embedded in a
medium, may also be described as the `momentum-independent self-energy
approximation'. It amounts to

(\textit{i}) replacing the general position-dependent self-energy $\Sigma
(R,\omega )$\textbf{\ }$\ $by the on-site value $\Sigma (0,\omega )$, and

(\textit{ii}) providing a prescription for computing $\Sigma (0,\omega )$
from the solution of a single-site quantum impurity model.

Considering the problem more generally, one may write $\Sigma (p,\omega )$
as an expansion in orthogonal functions and truncate the expansion at a
finite order, thereby replacing a general function by a small number of
frequency-dependent coefficients, which may be determined from the solution
of a several site \textit{fictive impurity model}: ``fictive'' because
although the several-site impurity model can be regarded as a cluster
self-consistently embedded in a medium, this cluster need not be a
sub-cluster of the physical lattice considered. It should be viewed as \
merely a device for computing the self-energy functions of interest. 
The general idea of taking an abstract view of the `impurity' or cluster
in the dynamical mean field method is not new. 
It is mentioned in Ref.~\onlinecite{Georges96}, and  has recently been 
elegantly exploited by Kotliar and co-workers to link band theory and the
dynamical mean field method.\cite{Savrasov01b} It is also the basis of an
approach to the practical solution of the single-site DMFT equations put
forward by Potthoff.\cite{Potthoff03} Here we apply the idea to the study of
spatial correlations. 

The idea that the cluster model is merely an algorithm for computing
coefficients in a low-order orthogonal function expansion of a physical self
energy gives insight into the `causality violations' which sometimes occur
in dynamical mean field schemes. In the context of cluster extensions of
dynamical mean field theory, the term `causality violation' means that for
some momenta and frequencies, the imaginary part of the approximate lattice
self-energy computed from the cluster approximation has the `wrong' sign. \
We shall argue that this is nothing but the `ringing' phenomenon familiar
from Fourier analysis: if one approximates a non-negative but sharply peaked
function by a few low order terms in an orthogonal function expansion, the
approximant will change sign. We shall use the analogy to propose 
several simple cures. We note that the `DCA' formalism used for example
in Refs.~\onlinecite{Hettler98,Moukouri00,Maier02} corresponds to a choice of
orthogonal functions which are everywhere non-negative but which have no
common support. In this sense, `ringing' is absent, so the method is plainly causal, 
but discontinuities in momentum space occur. 

The outline of this paper is as follows. In section II we outline the
formalism we shall use. In section III we present examples of the use of
several-site impurity models to calculate the self-energy of the two
dimensional Hubbard model. Section IV discusses `causality violations' and
possible cures. Section V is a conclusion.

\section{Formalism}

We study a model of electrons moving on a $d$-dimensional lattice with short
ranged hopping amplitudes (described by a quadratic Hamiltonian $H_{0}$) and
interactions (described by a Hamiltonian $H_{int}$). Physical properties of
the model may be derived from the general `Luttinger-Ward' expression for
the action, which we may write in terms of the exact Green function $\mathbf{%
G}$ of the model as%
\begin{equation}
S=Tr\ln \left(- \mathbf{G}\right) + \overline{\Phi}_{skel}\left[\mathbf{G}%
\right]   \label{S}
\end{equation}%
with $\overline{\Phi}_{skel}[\mathbf{G}]$ 
the sum (with appropriate symmetry factors) of all vacuum to vacuum
`skeleton' diagrams drawn with full Green functions ($\mathbf{G}$) and no
self-energy insertions. The electron self energy $\mathbf{\Sigma}$ may be
obtained via%
\begin{equation}
\mathbf{\Sigma} = \delta \overline{\Phi }_{skel}/\delta \mathbf{G} .
\label{sigmadef1}
\end{equation}

The functional $S$ is defined for any $G$ once the interactions ($H_{int}$)
are specified. The correct $G$ for a given `band structure' ($H_{0}$) is
determined from the equation 
\begin{equation}
-\left( \delta _{t}-H_{0}\right) ^{-1}\equiv \mathbf{G}_{0}^{-1}=\frac{%
\delta S}{\delta \mathbf{G}} .  \label{g0def}
\end{equation}

It is convenient for our purposes to eliminate $\mathbf{G}$ in favor of $%
\mathbf{\Sigma }$ via a Legendre transformation (this transformation is also
basic to the work of Potthof \cite{Potthoff03}) defining 
\begin{equation}
\Phi_{skel} [\mathbf{\Sigma} ]=\overline{\Phi}_{skel}-Tr\left( \mathbf{%
\Sigma G}\right) ;  \label{omegaskel}
\end{equation}%
Note that it follows from Eqs. (\ref{sigmadef1},\ref{omegaskel}) that 
\begin{equation}
\mathbf{G}=-\frac{\delta \Phi_{skel}}{\delta \mathbf{\Sigma }} ,  \label{gdef}
\end{equation}%
and that in this representation the theory is fixed by demanding that the $%
\mathbf{G[\Sigma ]}$ obtained from Eq. (\ref{gdef}) is identical to the $%
\mathbf{G}$ obtained via%
\begin{equation}
\mathbf{G[\Sigma ]}=\left( \mathbf{G}_{0}^{-1}-\mathbf{\Sigma }\right) ^{-1}, 
\label{gfinal}
\end{equation}%
i.e. by minimizing with respect to $\mathbf \Sigma$ the functional%
\begin{equation}
\Omega \lbrack \mathbf{\Sigma} \rbrack = -Tr \ln \left(-\mathbf{G}_{0}^{-1}+%
\mathbf{\Sigma }\right) +\Phi_{skel}[\mathbf{\Sigma}] .  \label{phifinal}
\end{equation}

The original momentum-independent-self-energy (single-site) dynamical mean
field approximation may be formulated from Eqs. (\ref{omegaskel},\ref{gfinal})
as follows (essentially this derivation is given in Ref.~\onlinecite{Georges96}): 
define $\Phi_{loc}$ 
as the approximation to the exact $\Phi_{skel}[\mathbf{\Sigma }(p,\omega) ]$
obtained by replacing the exact momentum dependent self-energy by a local
approximant $\mathbf{\Sigma }_{loc}(\omega )=\int (dp)\mathbf{\Sigma }%
(p,\omega )$ which depends only on frequency.\ Consistency demands that this
be equivalent to replacing $\overline{\Phi}
_{skel}\left[ \mathbf{G}\right] $ in Eq. (\ref{omegaskel}) by the analogous
quantity defined with the local Green function. Because $\mathbf{\Sigma }%
_{loc}(\omega )$ depends only on frequency, the quantity which follows from
Eq. (\ref{gdef}) is the local Green function $\mathbf{G}_{loc}=\int (dp)\mathbf{%
G}(p,\omega )$ and Eq. (\ref{gfinal}) becomes the relation $\mathbf{G}%
_{loc}=\int (dp)\left[ \mathbf{G}_{0}^{-1}(p,\omega )-\mathbf{\Sigma }%
_{loc}(\omega )\right] ^{-1}$.

The crucial observation which makes the \ single-site dynamical mean field
approximation useful is that because $\Phi_{loc}$ 
is a functional only of a function of frequency, it may be defined
nonperturbatively as the solution of a single-site (quantum-impurity) model
which is specified by a frequency-dependent Weiss field and by interaction
terms related to the local interactions of the original model. The Weiss
field is fixed by demanding that Eq. (\ref{gfinal}) is satisfied when the left
hand side of this equation is the Green function calculated from the
impurity model and the right hand side is the local Green function
calculated from the lattice Hamiltonian, using the impurity model
self-energy.

A generalization is now evident. Consider a set of functions $\{\phi
_{i},\psi _{i}\}$ such that 
\begin{equation}
\delta _{pp^{\prime }}=\sum_{i}\phi _{i}(p)\psi _{i}(p^{\prime }) , 
\label{complete}
\end{equation}%
so that 
\begin{equation}
\mathbf{\Sigma }(p,\omega )=\sum_{i}\phi _{i}(p)\mathbf{\Sigma }_{i}
(\omega )  \label{sigorthog}
\end{equation}%
with $\mathbf{\Sigma }_{i}(\omega )=\int \left( dp\right) \psi
_{i}(p)\mathbf{\Sigma }(p,\omega )$. Inserting Eq. (\ref{sigorthog}) into
Eq. (\ref{omegaskel}) yields%
\begin{equation}
\Phi_{skel}[\{\mathbf{\Sigma}_i \}]=\overline{\Phi}_{skel}-\sum_{i}Tr\left( 
\mathbf{\Sigma }_i \mathbf{G}_i \right)  \label{omegaskepprewrite}
\end{equation}%
with $\mathbf{G}_{i}=\int \left( dp\right) \phi _{i}(p)\mathbf{G}(p,\omega )$.

Now, construct an approximant for the self-energy as a sum of a small number
of terms in the expansion given in Eq.~(\ref{sigorthog}) 
\begin{equation}
\mathbf{\Sigma }(p,\omega )\approx \mathbf{\Sigma }_{approx}(p,\omega
)\equiv \sum_{i=0..n}\phi _{i}(p)\mathbf{\Sigma }_{i}(\omega ). 
\label{sigapprox}
\end{equation}%
Define $\Phi_{approx}[\mathbf{\Sigma }_{approx}]$ 
as the functional obtained from $\Phi_{skel}$ 
by using the approximate self-energy instead
of the exact $\mathbf \Sigma$. Eq (\ref{gdef}) implies 
that this construction is equivalent to approximating $\bar{\Phi}_{skel}$
by the set of diagrams drawn using only the $\mathbf G$ 
conjugate [in the sense defined below Eq.~(\ref{omegaskepprewrite})]
to the retained $\mathbf{\Sigma}_i$.
We see that $\Phi_{approx}$ 
is a functional of $n+1$ frequency dependent fields. It therefore
corresponds to the solution to some $\left( n+1\right) $-site \textit{%
fictive impurity model} involving $n+1$ Weiss fields, and interactions
derived (as discussed below) from the original model. The Weiss fields are
fixed by the requirement that the impurity-model Green functions $%
\mathbf{G}_{i}=-\delta \Phi_{approx}/\delta \mathbf{\Sigma}_{i}$ 
are given by appropriate integrals over the lattice Green function:%
\begin{equation}
\mathbf{G}_{i}(\omega )=\int (dp)\phi _{i}(p)
\left[ \mathbf{G}_{0}^{-1}(p,\omega )-\mathbf{\Sigma}_{approx}(p,\omega )\right] ^{-1}  . 
\label{sce}
\end{equation}%
From this point of view we may interpret the impurity model simply as a
mathematical means for calculating, nonperturbatively, an approximation to
the self-energy.

The original dynamical
mean field method corresponds to retaining only 
the $i=0$ term in the self energy.
In the ``DCA'' approach of Jarrell and co-workers,\cite{Hettler98} 
the functions $\phi _{i}$ 
are obtaining by tiling the Brillouin zone into regions $\mathcal{R}_{i}$
and setting $\phi _{i}(p)=1$ if $p$ is contained in $\mathcal{R}_{i}$ and $%
\phi _{i}(p)=0$ otherwise. These functions are clearly nonnegative
everywhere, and are orthogonal because they have no common support: at any $p
$ exactly one $\phi _{i}$ is non-zero. This choice of functions leads
approximants with discontinuities in momentum space. 
Other choices are discussed below. 

To completely specify the impurity model we must determine the interaction
terms. Specifying the interaction terms is a subtle issue
in general. One approach is to observe that
the skeleton functional, and therefore its
approximant $\Phi_{approx}$ 
is defined for any hopping Hamiltonian $H_{0}$. We may  therefore consider the
special case of no hopping (so $H_{0}$ is simply the energies of whatever
on-site levels are considered).  For models (such as the Hubbard model)
in which the interaction is local, 
both the Green function and self-energy are
diagonal in the site representation and are easy to compute. Comparison of
the exact and approximate expressions then shows (as was already
demonstrated for the single-site DMFT in Ref.~\onlinecite{Georges96}) that
the interaction terms in
the impurity model are simply the original interaction terms of the
lattice model. However, for longer ranged interactions, the situation
may become more complicated. Indeed the difficulty 
with longer range interactions appears to be common to
all schemes. For example, in real space cluster
schemes the issue of interactions connecting
the cluster to the medium must be addressed
\cite{Si96,Chitra01,Sun02} while in  the `DCA'  
the Laue-function arguments advanced by Aryanpour {\it et al}.\cite{Aryanpour03} 
require a momentum-independent interaction.
A separate paper will analyze the issue from the present
point of view.\cite{Okamoto03b}

\section{Example Approximants}

\subsection{General considerations}

A cluster extension of dynamical mean field theory involves finding an
impurity model to represent the frequency-dependent expansion coefficients
in Eq. (\ref{sigorthog}). Such a model must involve $n$ fields which have an
orthogonality property, so that we may unambiguously determine $n$
independent Green functions and self energies. One convenient choice is to
introduce an $m>n$ component spinor of fermions $\psi $ and write an action
of the form 
\begin{equation}
S=S^{(2)}+S_{int}  \label{S1}
\end{equation}%
with (we have suppressed the spin indices here) 
\begin{equation}
S^{(2)}=\psi ^{\dagger }\Biggl[\sum_{i=0,...,m}\!\!b_{i}\mathbf{M}_{i}\Biggr]%
\psi ,  \label{S2}
\end{equation}%
where $b_{i}$ are frequency dependent Weiss fields and the $m\times m$
matrices $\mathbf{M}_{i}$ satisfy 
\begin{equation}
Tr[\mathbf{M}_{i}\mathbf{M}_{j}]=m\delta _{ij} . \label{orthog}
\end{equation}

A general impurity model Green function is given by
\begin{equation}
\mathbf{G}_{imp}=g_{0} \mathbf{M}_{0}+g_{1} \mathbf{M}_{1}+g_{2} \mathbf{M}_{2}+...  
\label{gimpgen}
\end{equation}%
with coefficients $g_i(\omega)$ given by 
\begin{equation}
g_{i}=\frac{1}{m} \frac{\delta \ln Z_{imp}}{\delta b_{i}} . \label{gigen}
\end{equation}

The orthogonality relations imply the self energies
\begin{equation}
\Sigma _{imp,i}= \frac{1}{m} Tr\left[ \mathbf{M}_{i}\left(
\sum_{j}b_{j} \mathbf{M}_{j}- \mathbf{G}_{imp}^{-1}\right) \right]  \label{sigmai}
\end{equation}%
leading to self consistency equations of the form Eq. (\ref{sce}). The self
consistency condition in the local (no-hopping) limit implies $\mathbf M_0$ is the
unit matrix.

\subsection{Harmonic expansion--second order}

Here we write, for a hypercubic lattice in $d$ dimensions%
\begin{equation}
\Sigma _{approx}(p,\omega )=\Sigma _{0}(\omega )+\sum_{a}e^{ip\cdot a}\Sigma
_{a}(\omega )  \label{sigharm}
\end{equation}%
with $a=\pm x,\pm y...$ so that naively there are $2d+1$ mean field
equations:%
\begin{eqnarray}
G_{0} &=&\int \left( dp\right) G_{p}(\omega ) , \label{gharm0} \\
G_{a} &=&\int \left( dp\right) e^{ip\cdot a}G_{p}(\omega ) , \label{gharm1}
\end{eqnarray}%
so we should write an impurity model action involving $2d+1$ fields.
However, we may argue that in a cubic lattice, all of the components $\Sigma
_{a}$ are equal, so that the physics may be expressed via an impurity model
depending on two fields, $\Sigma _{0}$ and $\Sigma _{a}$. There are two self
consistency conditions. The impurity model is then specified by a partition
function $Z^{(2)}$ arising from a functional integral over the action%
\begin{equation}
S^{(2)}=%
\begin{pmatrix}
\psi _{1}^{\dag } & \psi _{2}^{\dag }%
\end{pmatrix}%
\begin{pmatrix}
b_{0} & b_{1} \\ 
b_{1} & b_{0}%
\end{pmatrix}%
\begin{pmatrix}
\psi _{1} \\ 
\psi _{2}%
\end{pmatrix}%
+S_{int}^{(2)} . \label{s2}
\end{equation}

The mean field equations may be written in symmetrized form as%
\begin{eqnarray}
G_{0,imp} &=&\frac{\delta \ln Z^{(2)}}{2\delta b_{0}}=\int \left( dp\right)
G_{p}(\omega ) , \label{g20} \\
G_{1,imp} &=&\frac{\delta \ln Z^{(2)}}{2\delta b_{1}}=\int \left( dp\right)
\gamma _{p}G_{p}(\omega ) , \label{G21}
\end{eqnarray}%
where 
\begin{equation}
G_{p}(\omega )=\frac{1}{\omega -\varepsilon _{p}-\Sigma _{0}-2d\gamma
_{p}\Sigma _{1}} . \label{Gp2}
\end{equation}

One may consider including longer ranged correlations--for example, in $d>1$
the second neighbor correlation, writing%
\begin{eqnarray}
\Sigma _{approx}(p,\omega ) &=&\Sigma _{0}(\omega )+\sum_{a}e^{ip\cdot
a}\Sigma _{1}(\omega )  \notag \\
&&+\sum_{a,b\neq a}e^{i\left( p\cdot a+p\cdot b\right) }\Sigma _{2}(\omega ). 
\label{sig2nd}
\end{eqnarray}%
We must then seek a multi-site impurity model which depends upon three Weiss
fields $b_{0}$, $b_{1}$, $b_{2}$ and involves three self-energy functions.
When written in matrix form the impurity model must thus involve a closed
algebra of three orthogonal symmetric matrices $\mathbf{M}_{0}$, $\mathbf{M}_{1}$, 
$\mathbf{M}_{2}$. We
have not found a suitable closed algebra of $3\times 3$ matrices; however a
closed algebra of $4\times 4$ matrices exists, with 
\begin{eqnarray}
\mathbf{M}_{0} &=&1 , \label{M0} \\
\mathbf{M}_{1} &=&\frac{1}{\sqrt{2}}%
\begin{pmatrix}
0 & 1 & 0 & 1 \\ 
1 & 0 & 1 & 0 \\ 
0 & 1 & 0 & 1 \\ 
1 & 0 & 1 & 0%
\end{pmatrix} ,
\label{M1} \\
\mathbf{M}_{2} &=&%
\begin{pmatrix}
0 & 0 & 1 & 0 \\ 
0 & 0 & 0 & 1 \\ 
1 & 0 & 0 & 0 \\ 
0 & 1 & 0 & 0%
\end{pmatrix} ,
\label{M2}
\end{eqnarray}%
This choice corresponds to a four-site real-space cluster with the same topology as
that considered by Lichtenstein and Katsnelson in a pioneering study of
superconductivity and antiferromagnetism in the Hubbard model.\cite%
{Lichtenstein00} 

The impurity model action becomes%
\begin{equation}
S^{(4)}=Tr\ln \left[ \sum_{\alpha =0,1,2}b_{\alpha } \mathbf{M}_{\alpha }\right]
+ S^{(4)}_{int}
\end{equation}%
leading to the self consistency conditions 
\begin{equation}
\frac{1}{4} \frac{\delta \ln Z^{((4)}}{\delta b_{\alpha }}=\int
\left( dp\right) \phi _{\alpha }(p)G(p,\omega )  \label{sce3}
\end{equation}%
with ($z$ is the number of nearest neighbors.) 
\begin{eqnarray}
\phi _{0} &=&1 , \label{phi0} \\
\phi _{1} &=&e^{ip\cdot a}=\frac{1}{z}\sum_{a}e^{ip\cdot a} , \label{phi1} \\
\phi _{2} &=&e^{i\left( p\cdot a+p\cdot b\right) }=\frac{1}{z(z-1)}%
\sum_{a,b}e^{i\left( p\cdot a+p\cdot b\right) } . \label{phi2}
\end{eqnarray}

We finally discuss the interaction terms. 
In models with purely on-site interactions, the ${\cal S}^{(n)}_{int}$ may be fixed by 
analysis of the local limit, in which $G_\alpha = 0$ for $\alpha \ne 0$ and $G_0$ is known. 
For example, in the Hubbard model, $G_{0}$ has poles at $\omega=\pm U/2$. 
The absence of any intersite 
correlations ensures that any off-diagonal terms in $S_{int}^{(n)}=0$ and the
identity of the $n$ sites implies the interaction is just%
\begin{equation}
S_{int}^{(n)}=U\sum_{j=1,...,n}n_{j\uparrow }n_{j\downarrow } .
\label{S2int}
\end{equation}

\begin{figure}[tbp]
\epsfxsize=1\columnwidth \centerline{\epsffile{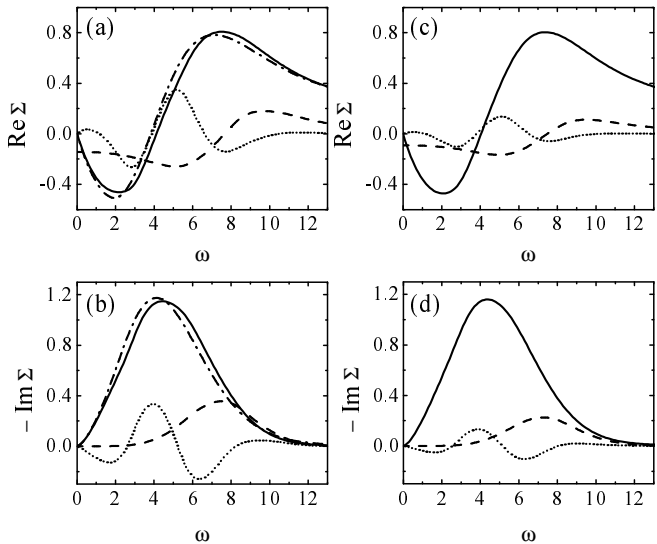}}
\caption{(a) and (b): Real and imaginary part of self-energy obtained from
`iterated perturbation theory' approximation to half-filled Hubbard model ($%
U=4, T=0$) using one and four-site clusters. Dash-dotted curves are for
one-site cluster. Solid, dashed, and dotted curves are $\Sigma_0$, $\Sigma_1$%
, and $\Sigma_2$ in four-site clusters, respectively. $\Sigma_1$ and $%
\Sigma_2$ are multiplied by 4. (c) and (d): Self energies in four-site
clusters obtained from simple filtering procedure of section IV, with $%
f_1=0.85$ and $f_2=f_1^2$. }
\label{fig:ipt}
\end{figure}

When treated perturbatively to order $U^{2}$, the two and four impurity
models reproduce exactly the appropriate Fourier coefficients of the exact
(perturbative) lattice self-energy of the Hubbard model. 
To gain some idea of effects beyond
perturbation theory we show in panels (a,b) Fig. \ref{fig:ipt} results for
the real and imaginary parts of the self-energy obtained from the `IPT'
approximation,\cite{Georges96} in which the impurity model and self
consistency condition are solved by writing the second order perturbation
theory expression for the self-energy, but using exact (impurity model)
Green functions. We observe that in the IPT approximation to the Hubbard
model, non local effects are very small.

\section{Causality}

\subsection{General considerations and a simple example}

A key difficulty in `cluster' extensions of the coherent potential
approximation or of dynamical mean field theory has been causality. Any
physical theory must be causal, which implies in particular that 
$Im \mathbf{\Sigma }(p,\omega )<0$. Many cluster schemes however generate
functions $\mathbf{\Sigma }_{approx}(p,\omega )$ with the unfortunate
property that for some range of $p,\omega $, 
$Im \mathbf{\Sigma }_{approx}(p,\omega )>0.$ These violations of causality
have apparently never been clearly understood or cured, but are generally
viewed a deficiencies of the `cluster model' used to calculate the
self-energy.

The formal development of the previous section suggests that the causality
violations may be thought of as an example of the `ringing' phenomenon
familiar from Fourier analysis. Any straightforward real-space cluster
scheme corresponds to using some set of orthogonal functions to expand \ the
momentum dependence of the exact lattice self-energy $\mathbf{\Sigma }%
(p,\omega )$ in the sense of Eq. (\ref{sigapprox}). The first term in an
orthogonal function expansion is $\phi _{0}=1$ which is everywhere positive,
so the local approximation is guaranteed to be causal, but all of the other
orthogonal functions change sign over the Brillouin zone, so an expansion
which is truncated at low order is not guaranteed to be positive everywhere
in the zone. In particular, if at fixed $\omega $ the function $\mathbf{%
\Sigma }(p,\omega )$ has a strong, narrow peak at some momentum $p$, then
truncating an orthogonal function expansion at a low order will produce a
self-energy whose imaginary part changes sign. As noted above, in the ``DCA''
approach \cite{Hettler98} this problem is avoided, at the expense of
introducing discontinuities in the momentum space representations of $G$ and $%
\Sigma$, by choosing the orthogonal functions to be a tiling of the
Brillouin zone by rectangular filters. However, the general argument, that
acausal behavior found in most schemes arises from `ringing' associated with
narrow peaks in the momentum dependent self-energy, suggests also that the
phenomenon is not of very great significance.

\begin{figure}[tbp]
\epsfxsize=0.8\columnwidth \centerline{\epsffile{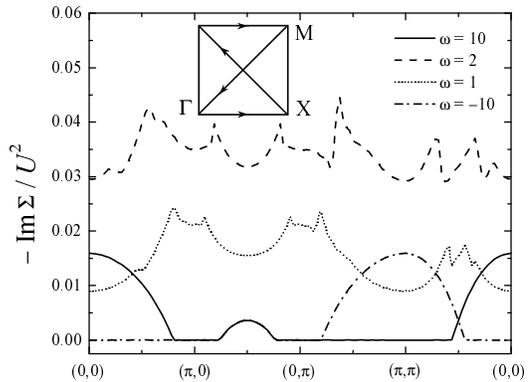}}
\caption{Momentum dependence of the second order perturbation theory
approximation to the imaginary part of the self-energy of the half-filled
two dimensional Hubbard model at $T=0$. Solid, dashed, dotted and
dash-dotted curves are for $\protect\omega = 10, 2,1$ and $-10$,
respectively.}
\label{fig:exact}
\end{figure}

To illustrate the issue we show in Fig. \ref{fig:exact} the momentum
dependence of the second order perturbation theory approximation to the
imaginary part of the self-energy of the two dimensional square-lattice
Hubbard model, for several different frequencies. 
At frequencies within the electronic band the
momentum dependence of the self-energy is relatively weak, but at
frequencies well above the upper band edge or well below the lower band
edge, the imaginary part of the self-energy becomes sharply peaked in
momentum space. This phenomenon has a simple kinematic origin. At this order
of perturbation theory, the imaginary part of the self-energy at $\omega > 0$
corresponds to a decay of a particle into two particles and a hole. At $%
\omega = 10$, energy conservation means the allowed final states correspond
to two particles near the top of the band [momenta near $(\pi,\pi)$] and a
hole near the bottom of the band [momenta near $(0,0)$]. Momentum
conservation then restricts the initial momentum to be near $(0,0)$.

We now consider expanding the self-energy in Fourier harmonics. The symmetry
of the hypercubic lattice implies that 
\begin{equation}
\Sigma (p,\omega )=\Sigma _{0}(\omega )+2d\gamma _{p}\Sigma _{1}(\omega
)+2d\left( d-1\right) \gamma _{p}^{(2)}\Sigma _{2}(\omega )+...
\label{sigexp}
\end{equation}%
with, for a hypercubic lattice of unit lattice constant in $d$ dimensions, 
\begin{eqnarray}
\gamma _{p} &=&\frac{1}{d}\sum_{a=1...d}\cos (p_{a}) , \label{gammadef} \\
\gamma _{p}^{(2)} &=&\frac{1}{d\left( d-1\right) }\sum_{\substack{ a=1...d 
\\ b\neq a}}\cos (p_{a})\cos (p_{b}) . \label{gamma2def}
\end{eqnarray}

\begin{figure}[tbp]
\epsfxsize=0.8\columnwidth \centerline{\epsffile{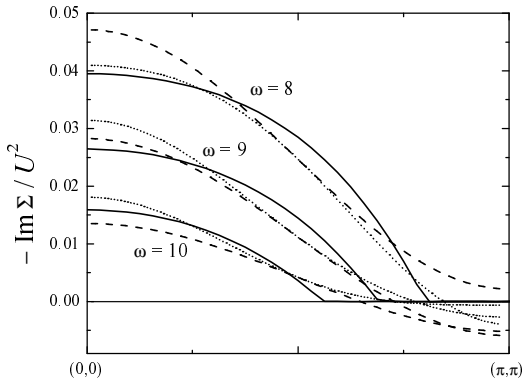}}
\caption{Comparison of momentum dependence of exact second order
perturbation theory approximation to self-energy of half-filled two
dimensional Hubbard model (solid line) and the approximation using Eq. 
(\protect\ref{sigexp}) at $T=0$; dash and dotted curves are the approximation
using the first and second, and first three terms in Eq. (\protect\ref{sigexp})%
. }
\label{fig:approx}
\end{figure}

The dashed and dotted curves in Fig. \ref{fig:approx} show the result of
approximating the exactly calculated $\Sigma $ by the first and second, or
first three terms in the series given in Eq. (\ref{sigexp}). An acausal
behavior is observed at frequencies well above the upper band edge or well
below the lower band edge, in accordance with the qualitative arguments
presented above. At $\omega =8$, we observe the acausal behavior does not
occur if only the first neighbor term is retained (dashed line) but does
occur if both first and second neighbor terms are retained.  This is because 
$Im\Sigma _{2}(\omega )$ changes sign between $\omega =8$ and 9 to fit the
exact $Im\Sigma (p,\omega )$. [One can see this behavior even in Fig. \ref%
{fig:ipt} (b,d).] 

\begin{figure}[tbp]
\epsfxsize=0.8\columnwidth \centerline{\epsffile{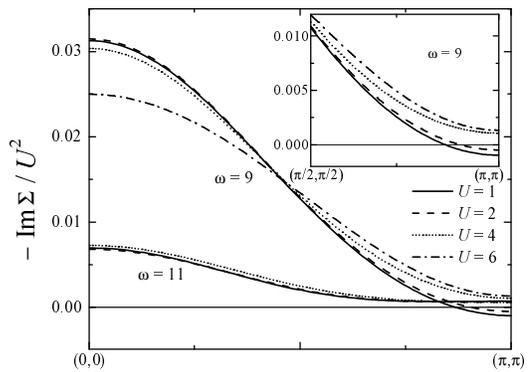}}
\caption{Main panel: momentum dependence of $Im$ $\Sigma(\omega)$ at $\omega=9,11$ calculated 
for  $U=1,2,4,6$ by use of the `IPT' approximation on a four-site cluster. The increase of
$U$ is seen to lead to a broadening of the peak, which for the large $U$ values is seen (inset)
to eliminate the `acausal' behavior  entirely. }
\label{fig:iptlargeu}
\end{figure}

As noted above the acausal behavior has (at least in this instance) a simple
kinematic origin, which suggest that it may not occur in all circumstances.
For example, at larger $U$, more complicated decay channels (e.g. one
particle decaying into 3 particles and 2 holes) with fewer kinematic
constraints may become important, leading to a broadening of peaks in the
momentum dependent self energy and therefore to a causal self energy. An
example of this phenomenon is shown in Fig \ref{fig:iptlargeu}, which displays
the imaginary part of the self energy calculated  high frequencies in the
`IPT' approximation at several different $U$-values. \ The broadening of the
peak and the destruction of the acausal behavior suggested by the arguments
above are clearly observed. We stress that the IPT is an uncontrolled
approximation---however it is a computationally tractable example
illustrating a phenomenon which we suspect is of more general significance.

\subsection{Filtering}

We have argued that any cluster extension of the dynamical mean field method
amounts to a scheme for computing the coefficients in a low-order orthogonal
function expansion of the electronic self-energy. This suggests that the
reported causality difficulties are generic and inescapable: \textit{no}
truncation of an orthogonal function expansion is guaranteed to lead to an
everywhere non-negative approximant for an arbitrary test function. We
emphasize, however, that the impurity model itself \ is causal; it is only
the resulting lattice self-energy which may have acausal features.

The discussion of the previous subsection suggests that the acausal behavior
is `ringing' is associated with a relatively large peak near a particular
momentum. It is possible that such a peak could be physically important,
occurring for example at a fermi surface `hot spot' in a system close to a
quantum critical point. In such a case, long ranged interactions are
evidently physically crucial, and modeling them with a local model is
simply inappropriate. However, as seen in the explicit calculations present
above, peaks in 
$Im \Sigma $ which are both sharp and large relative to typical values of $%
\Sigma $ are more often associated with band edges and extremal frequencies.
These regions of $\omega $ and $p$ are not particularly important for
energetics, suggesting that the acausality is simply a minor technical
annoyance. In what follows we consider methods of removing it.

We first note that in particular temperature regimes of particular problems
(for example, the double exchange model on the simple cubic lattice and at
not too low temperatures \cite{Soloviev02}) the ringing phenomenon might not
occur. We also note that a clever choice of expansion functions may mitigate
the severity of the problem. For example, we saw in the Hubbard model the
difficulties arise from narrow peaks centered near the band edges $(p\approx
0$ and $p\approx (\pi ,\pi ,..)$). An expansion based on the functions $%
1,\gamma _{p}$ and a function orthogonal to both $1$ and $\gamma _{p}$ but
strongly peaked near $0$ and $(\pi ,\pi ,..)$ might have a wider range of
applicability than the straightforward harmonic expansion.

A more general approach is to filter the self-energy, for example by
convolving it with a function to smooth out any sharp peaks and then
approximating the smoothed function by a low order harmonic expansion. Let
us make this approach more precise, writing Eq. (\ref{sigapprox}) as%
\begin{equation}
\Sigma_{approx}(p,\omega )\equiv \sum_{i=0..n}f_{i}\phi _{i}(p)%
\Sigma_{i}(\omega ) , \label{sigsmooth}
\end{equation}%
where $f_{0}=1$ and $1>f_{i>0}$ $\succeq 0$ are the Fourier components of
the smoothing function. Carrying through the development of the previous
section leads to 
\begin{equation}
G_{i}(\omega )=\frac{\delta \Omega _{approx}}{\delta \Sigma _{i}(\omega )}%
=f_{i}\int \left( dp\right) G_{p}(\omega )\phi _{i}(p)  \label{gsmoothed}
\end{equation}%
so that the quantum impurity model acts to reproduce the smoothed Green's
functions of the theory. (We note that sum rules typically constrain the
large $\omega $ behavior of the local Green function, so that one must
choose $f_{0}=1$ in order to have a consistent representation at least
within the simple impurity models of which we have studied). Fig. \ref%
{fig:imp10} shows the results obtained using different choices of simple
filtering coefficients in the low order self energies discussed in the
previous section.

\begin{figure}[tbp]
\epsfxsize=0.8\columnwidth \centerline{\epsffile{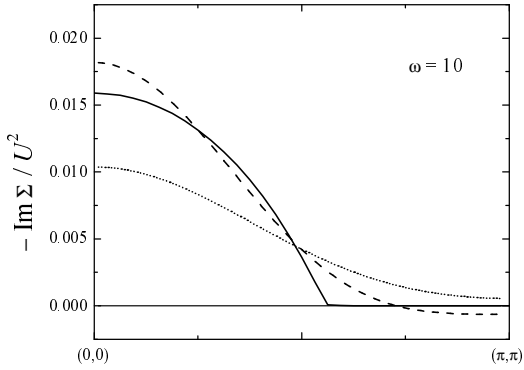}}
\caption{Comparison of filtered, unfiltered and exact results for second
order perturbation approximation to self-energy of the half-filled two
dimensional Hubbard model at $\protect\omega = 10$. Solid curve is the exact
result. Dashed and dotted curves are obtained from four-impurity model with
filtering coefficients $f_1 = 1$ and 0.85, respectively, and $f_2=f_1^2$.
Numerical calculation gives the following values for $Im \Sigma_i$: $(Im
\Sigma_0, Im \Sigma_1, Im \Sigma_2) = (-0.00418, -0.00234, -0.00114)$ for $%
f_1=1$ and $(-0.00418, -0.00144, -0.00033)$ for $f_1=0.85$. Calculations are
done at $T=0$. }
\label{fig:imp10}
\end{figure}

The straightforward filtering approach gives up some fraction of the
intersite correlations in order to guarantee a causal theory. A more
sophisticated possibility would be a frequency dependent filtering. We saw
from the low order perturbation calculation on the Hubbard model that the
self-energy was only very strongly peaked in momentum space for very high or
very low energy states, where decay kinematics constrained all states
involved to be near the band edges. These states are not very important to
energetics. One may therefore filter out only these by setting an arbitrary 
frequency scale $\omega_{f}$ and $\Delta$ and writing e.g. $f_{1}=1-\bar f /
[ exp \{(-|\omega|+\omega_f)/\Delta \} +1 ]$. We note that this filtering
may be performed \textit{ex-post-facto:} one may solve the impurity model,
determine the band-edge regions where filtering is required, and then
re-solve the impurity model with filtering only in these regions. Fig. \ref%
{fig:filter} shows the results obtained using the above frequency-dependent
filtering with different choices of parameters. Here, we take $f_2 = f_1^2$,
and four-impurity model is self-consistently solved. It is clearly shown
that $\Sigma_{1,2}$ are reduced at higher frequency region where acausal
behavior has been observed in the second order perturbation theory
approximation, while there is not much effect in the low frequency region.

\begin{figure}[tbp]
\epsfxsize=0.8\columnwidth \centerline{\epsffile{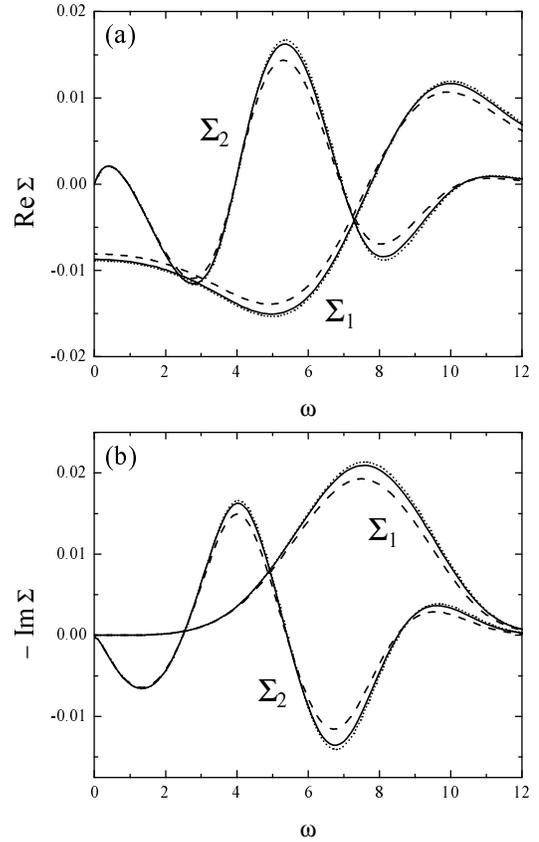}}
\caption{Real (a) and imaginary (b) part of self-energy, $\Sigma_1$ and $%
\Sigma_2$, obtained from `iterated perturbation theory' approximation to
half-filled Hubbard model ($U=2, T=0$) using four-site cluster with
frequency-dependent filtering. Solid and dashed curves are the results with
parameters $(\bar f, \protect\omega_f, \Delta) = (0.15, 6, 1)$ and (0.15, 4,
1), respectively. Dotted curves are the results without filtering. }
\label{fig:filter}
\end{figure}

A still more sophisticated approach would be to mix different harmonic
components, so that low order terms were modified only if higher order terms
were important (i.e. if filtering were really needed). This approach is most
helpful if a-priori knowledge of the locations of sharp momentum-space
structures is available. Indeed, the calculations presented above of the
behavior of the Hubbard model strongly suggest that the singularities are
associated with the top and bottom of the band, i.e. with states around $%
p=(0,0,..)$ or $p=(\pi ,\pi ,..)$. To define this transformation we begin
from the harmonic expansion, Eq. (\ref{sigapprox}), and then introduce a
transformation which mixes the harmonic coefficients, so that%
\begin{equation}
\Sigma_{approx}(p,\omega )\equiv \sum_{i,j=0..n}\phi _{i}(p)f_{ij}%
\Sigma_{j}(\omega ) , \label{sigsmoothed2}
\end{equation}%
and the self consistency equation becomes%
\begin{equation}
G_{i}(\omega )=\sum_{j}f_{ij}\int \left( dp\right) G_{p}(\omega )\phi _{j}(p) .
\label{gsmoothed2}
\end{equation}

We have however not yet explored these more complicated filtering procedures

\section{Conclusion}

In this paper we have argued that one should take a more abstract view of
cluster extensions of the dynamical mean field method; regarding them as
algorithms for computing frequency dependent coefficients in an orthogonal
function expansion of the electron self-energy. This approach renders moot
the debates about `correct' choice of cluster and embedding, and clarifies
the meaning of the `causality violation' encountered in real-space-cluster
extensions of the dynamical mean field method. We have suggested that the
causality violation is in most cases a minor technical problem which can be
cured \ if needed by any of a variety of `filtering' procedures.

Several questions arise, for which further research would be desirable. One
is the question of the correct choice of interaction terms in the ``fictive
impurity model''. 
This issue has been discussed in the context of 
on-site interactions, especially in connection with the ``DCA'',\cite{Aryanpour03} 
and, for longer ranged interactions
in the context of the E-DMFT approach
\cite{Chitra01,Sun02} but deserves more attention in the context of the more
conventional orthogonal function expansions.  
A second key issue is the origin of the `causality violations'.
We have shown that in simple perturbative models (chosen
because exact expressions for the momentum dependent self energy 
are available) the violations are an example of the
familiar `ringing' phenomenon, and moreover occur
mainly in band edge regions of little kinematical
importance. The issue however deserves further exploration
in less trivial contexts. A third open problem is the question of
which approximants to the self-energy are representable by impurity models.
We have presented arguments (substantiated by low-order perturbative
calculations in different limits) indicating that straightforward orthogonal
function expansions, with and without filtering, are representable. However,
one might imagine that more complicated approximate representations of the
self-energy might be advantageous in some problems. We do not at present
have a general framework for determining the circumstances under which a
general approximant to $\Sigma $ is representable in terms of an impurity
model. The issue of different choices of filtering function seems also
likely to benefit from further research. In particular, it seems likely that
one can more efficiently exploit the observation that the acausality is
associated mainly with the $p=0$ and $p=(\pi ,\pi ,...)$. Finally, applying
the method to a wider range of models, to explore which choices lead to good
representations of the physics of interest, is an urgent task.

\textit{Acknowledgements} We acknowledge very helpful conversations with B.
G. Kotliar and J. Serene. This research was supported by NSF DMR-00081075,
the DAAD and the CNRS, and DFG SFB608. AJM acknowledges the hospitality of the Bonn
University physics department and the ESPCI, and HM the hospitality of
Columbia University. SO acknowledges the financial support of JSPS.


\bigskip

\begin{thebibliography}{99}
\bibitem{Georges96} A.Georges, B. G. Kotliar, W. Krauth and M. J. Rozenberg,
Rev. Mod. Phys., \textbf{68}, 13 (1996).

\bibitem{Metzner88} W. Metzner and D. Vollhardt, Phys. Rev. Lett.
{\bf 62} 324 (1989); E. Mueller-Hartmann, Z. Phys. {\bf 74} 507 (1989)
and U. Brandt and C. Mielsch, Z. Phys. {\bf 75} 365 (1989).

\bibitem{Rozenberg95} M. J. Rozenberg, G. Kotliar, and H. Kajueter, G. A.
Thomas and D. H. Rapkine, J. M. Honig and P. Metcalf, Phys. Rev. Lett. 
\textbf{75}, 105-8 (1995).

\bibitem{Millis96b} A. J. Millis, R. Mueller and B. I. Shraiman, Phys. Rev 
\textbf{B54} 5405 (1996).

\bibitem{Savrasov01a} S Savrasov, B. G. Kotliar and Elihu Abrahams, Nature 
\textbf{410} 793 (2001).

\bibitem{Savrasov01b} S. Savrasov and B. G. Kotliar, pps 259-301 in \textit{%
New Theoretical Approaches to Strongly Correlated Systems}, A.M. Tsvelik
Ed., (Kluwer Academic Publishers: 2001).

\bibitem{Held01} K. Held, I.A. Nekrasov, G. Keller, V. Eyert, N. Bl\"{u}mer,
A.K. McMahan, R.T. Scalettar, T. Pruschke, V.I. Anisimov, D. Vollhardt,
cond-mat/0112079, to appear in \textit{Proceedings of the Winter School on
``Quantum Simulations of Complex Many-Body Systems: From Theory to Algorithms''%
}, February 25 - March 1, 2002,

\bibitem{Hettler98} M. H. Hettler, A. N. Tahvildar-Zadeh, M. Jarrell, T.
Pruschke, H. R. Krishnamurthy, Phys. Rev. \textbf{B58}, 7475-9 (1998).

\bibitem{Moukouri00} S. Moukouri, C. Huscraft and M. Jarrell,
(cond-mat/0004279).

\bibitem{Lichtenstein00} A. I. Lichtenstein and M. I. Katsnelson, Phys. Rev. 
\textbf{B62}, 9283-6 (2000).

\bibitem{Kotliar02} B. G. Kotliar, S. Y. Savrasov, G. Palsson and G. Biroli, Phys.
Rev. Lett. \textbf{87} 186401/1-4 (2001).

\bibitem{Biroli02} G. Biroli and B. G. Kotliar, Phys. Rev \textbf{B65}
155112 (2002)

\bibitem{Bolech02} C. Bolech, S. S. Kancharla and B. G. Kotliar,
cond-mat/0206166.

\bibitem{Aryanpour03} K. Aryanpour, M. H. Hettler, and M. Jarrell Phys. Rev. 
\textbf{B 67}, 085101 (2003).

\bibitem{DuCastelle75} F. DuCastelle, J. Phys. \textbf{C8} 3297 (1975).

\bibitem{Soloviev02} I Soloviev, cond-mat/0207544.

\bibitem{Si96} Q. Si and J. L. Smith, 
Phys. Rev. Lett. \textbf{77}, 3391 (1996).

\bibitem{Chitra01} R. Chitra and B. G. Kotliar, Phys. Rev. \textbf{B63},
115110 (2001).

\bibitem{Sun02} P. Sun and B. G. Kotliar, Phys. Rev. \textbf{B66}, 85120
(2002).

\bibitem{Mazurenko02} V. V. Mazurenko, A. I. Lichtenstein, M. I. Katsnelson,
I. Dasgupta, T. Saha-Dasgupta, and V. I. Anisimov Phys. Rev. \textbf{B 66},
081104 (2002)

\bibitem{Maier02} Th.A. Maier, M. Jarrell, A. Macridin, F.-C. Zhang,
unpublished (cond-mat/0208419).

\bibitem{Potthoff03} M. Potthoff, cond-mat/0301137.

\bibitem{Okamoto03b} S. Okamoto and A. J. Millis, in preparation.
\end{thebibliography}
\end{document}